\documentclass[conference]{IEEEtran}
\usepackage{subfigure}
\usepackage{setspace}
\usepackage{amsmath}
\usepackage{amssymb}
\usepackage{amsfonts}
\usepackage{amscd}
\usepackage{mathrsfs}
\usepackage[final]{graphicx}
\usepackage{graphicx}
\usepackage{psfrag}
\usepackage{color}
\usepackage{url}

\ifCLASSINFOpdf
\else
\fi
\begin{document}
%
% paper title
% can use linebreaks \\ within to get better formatting as desired
\title{The Fourier Transform Approach to Quantum Coding}

% author names and affiliations
% use a multiple column layout for up to three different
% affiliations
\author{
\IEEEauthorblockN{Hari Dilip Kumar}
\IEEEauthorblockA{Dept. of ECE,\\ Indian Institute of Science \\
Bangalore 560012, India\\
{\em Email:} \url{hari.coding@gmail.com}\\
}
\and
\IEEEauthorblockN{B. Sundar Rajan}
\IEEEauthorblockA{Dept. of ECE, \\Indian Institute of Science \\
Bangalore 560012, India\\
%\IEEEauthorblockA{\IEEEauthorrefmark{1}Corresponding author\\
\hspace{0.2cm} {\em Email:} \url{bsrajan@ece.iisc.ernet.in}\\
}
}

% conference papers do not typically use \thanks and this command
% is locked out in conference mode. If really needed, such as for
% the acknowledgment of grants, issue a \IEEEoverridecommandlockouts
% after \documentclass

% for over three affiliations, or if they all won't fit within the width
% of the page, use this alternative format:
% 
%\author{\IEEEauthorblockN{Michael Shell\IEEEauthorrefmark{1},
%Homer Simpson\IEEEauthorrefmark{2},
%James Kirk\IEEEauthorrefmark{3}, 
%Montgomery Scott\IEEEauthorrefmark{3} and
%Eldon Tyrell\IEEEauthorrefmark{4}}
%\IEEEauthorblockA{\IEEEauthorrefmark{1}School of Electrical and Computer Engineering\\
%Georgia Institute of Technology,
%Atlanta, Georgia 30332--0250\\ Email: see http://www.michaelshell.org/contact.html}
%\IEEEauthorblockA{\IEEEauthorrefmark{2}Twentieth Century Fox, Springfield, USA\\
%Email: homer@thesimpsons.com}
%\IEEEauthorblockA{\IEEEauthorrefmark{3}Starfleet Academy, San Francisco, California 96678-2391\\
%Telephone: (800) 555--1212, Fax: (888) 555--1212}
%\IEEEauthorblockA{\IEEEauthorrefmark{4}Tyrell Inc., 123 Replicant Street, Los Angeles, California 90210--4321}}

% use for special paper notices
%\IEEEspecialpapernotice{(Invited Paper)}

% make the title area
\maketitle

\begin{abstract}
%\boldmath
Quantum codes are subspaces of the state space of a quantum system that are used to protect quantum information. Some common classes of quantum codes are stabilizer (or additive) codes, non-stabilizer (or non-additive) codes obtained from stabilizer codes, and Clifford codes. We analyze these in the framework of the Fourier inversion formula on a group algebra, the group in question being a subgroup of the error group considered. We study other possible code spaces that may be obtained via such an approach, obtaining codes that are the direct sums of translates of Clifford codes, and more general codes obtained from idempotents in the transform domain. We derive necessary and sufficient conditions for error detection by direct sums of translates of Clifford codes, and provide an example using an error group with non-Abelian index group.
\end{abstract}
% IEEEtran.cls defaults to using nonbold math in the Abstract.
% This preserves the distinction between vectors and scalars. However,
% if the conference you are submitting to favors bold math in the abstract,
% then you can use LaTeX's standard command \boldmath at the very start
% of the abstract to achieve this. Many IEEE journals/conferences frown on
% math in the abstract anyway.

% no keywords

% For peer review papers, you can put extra information on the cover
% page as needed:
% \ifCLASSOPTIONpeerreview
% \begin{center} \bfseries EDICS Category: 3-BBND \end{center}
% \fi
%
% For peerreview papers, this IEEEtran command inserts a page break and
% creates the second title. It will be ignored for other modes.
\IEEEpeerreviewmaketitle

\section{Introduction}
% no \IEEEPARstart
% You must have at least 2 lines in the paragraph with the drop letter
% (should never be an issue)

Quantum error control codes allow to protect the computational state of a quantum computer against decoherence errors \cite{beyond2}. The well-known classes of quantum codes, namely stabilizer or additive codes \cite{gottes,crss}, non-additive codes (obtained as sums of translates of additive codes \cite{rains2, krp}) and Clifford codes \cite{Knill, clifford, beyond2}, are constructed as subspaces of the Hilbert space of the quantum system under consideration. Another type of quantum coding is subsystem coding \cite{subsystem}, which we do not consider in this paper. Stabilizer codes are obtained in \cite{gottes} as the joint +1 eigenspace of an Abelian subgroup of the error group on $n$ qubits. They are obtained in \cite{crss} as a joint eigenspace of a normal Abelian subgroup of the error group on $n$ qubits. Such codes have been generalized to error groups on higher dimensional systems (non-binary qudits) \cite{ rains, nonbinary}.  The first true ``non-additive'' code with nontrivial minimum distance, a $((5,6,2))$ code was discovered in \cite{rains2}. This code turned out to be the direct sum of translates of a $[[5,0,3]]$ stabilizer code.
Since its discovery, other such nonadditive codes have been constructed \cite{non1, non2, non3}. In \cite{krp}, a framework for non-additive code construction based on the inverse Fourier transform was developed, and some codes were constructed. Another class of codes, the Clifford codes, are obtained from normal subgroups of error groups, via images of the projector formulae (\cite{beyond2}, Theorem 1). We study these classes of codes in the context of the inverse Fourier transform. The contributions of this paper are:\begin{itemize}
\item The most commonly used quantum codes are of two kinds: subspace codes \cite{crss} and subsystem codes \cite{subsystem}. In this paper, we classify all subspace codes using the Fourier inversion approach.
\item We classify the error correcting properties of the direct-sums of translates of Clifford codes using the Fourier transform approach
\item The codes which have been commonly studied are all derivable from Clifford codes or direct sums of translates of Clifford codes. We show that this is equivalent to codes obtainable from Fourier inversion using transform domain values that are either zero or identity. We try to study codes that are obtainable via ``non-invertible idempotents'' in the transform domain; these have not been studied before.
\end{itemize}

The remaining content of this paper is organised as follows:
Section II introduces provides motivation for using introducing the Fourier transform in quantum coding. Section III (The Fourier Inversion Formula and Classes of Quantum Codes) introduces the Fourier transform and the inverse Fourier transform on a group algebra. This is related to some classes of quantum codes. In Section IV (Sums of Translates of Clifford Codes), we derive the necessary and sufficient conditions for error detection for a class of codes that are the direct sums of translates of Clifford codes. In Section V, we present an example code for a four-dimensional coding system (qudit.)

The following notation is used in this paper:
$\mathbb{C}$ denotes the complex numbers. $d$ denotes the number of levels of each qudit. The state space of this qudit, $\mathbb{C}^{d}$, is denoted by $H$. $n$ denotes the number of qudits in the system under consideration, $U(d)$ denotes the group of unitary matrices of side $d\times d$. $E$ denotes the error group of $n$ qudits under consideration, and $\mathbb{C}S$ denotes the group algebra of sums $\left\lbrace \sum_{s\in S} T_{s}s| T_{s}\in \mathbb{C} \right\rbrace$ for a group $S$ in $E$. The set of irreducible representations of $S$ over $\mathbb{C}$, upto isomorphism, is denoted by $R=\left\lbrace \rho_{i}|i\in\left\lbrace 1,\ldots ,k \right\rbrace \right\rbrace $, and the set of irreducible characters of the group $S$ is denoted by Irr$(S)=\left\lbrace \chi_{\rho_{i}}|i\in\left\lbrace 1,\ldots k\right\rbrace  \right\rbrace $. Given a character $\chi$ of $S$, we form the character $\chi^{g}$ of $S$ by conjugation: $\chi^{g}(s) = \chi (gsg^{-1}) \forall s \in S$, for a given $g \in E$. The group of integers modulo $n$ under addition is denoted by $Z_{n}$. The group of automorphisms of a complex vector space $W$ is denoted by $GL(W)$. We denote the $2\times 2$ zero matrix by $0_{2}$ and the $2\times 2$ identity matrix by $I_{2}$ when required. $X, Y$ and $Z$ refer to the Pauli operators $ \left( \begin{array}{cc}
0 & 1 \\ 
1 & 0
\end{array} \right) $, $\left( \begin{array}{cc}
0 & -1 \\ 
1 & 0
\end{array} \right) $ and $\left( \begin{array}{cc}
1 & 0 \\ 
0 & -1
\end{array} \right) $
respectively. $P_{X_{+}}$ is the one-dimensional projector onto the +1 eigenspace of $X$. $P_{Y_{-}}$ is the one-dimensional projector onto the $-i$ eigenspace of $Y$, etc.

\section{Background and Motivation}

Consider a quantum system with $N$ levels. Let $G$ be a group of order $N^{2}$, with identity element $1$. A nice error basis \cite{beyond1} on $H=\mathbb{C}^{N}$ is a set $\left\lbrace \rho(g) \in U(N)|g\in G \right\rbrace $ such that:
(i) $\rho(1)$ is the identity matrix
(ii) Trace$(\rho(g))$ = $n\delta_{g,1} \forall g \in G$
(iii) $\rho(g)\rho(h)=\omega(g,h) \rho(gh)$ $ \forall g,h \in G$ with $\omega(g,h) \in \mathbb{C}$. The group closure  under multiplication of the nice error basis (alternatively, the $\omega$-covering of the nice error basis) is called the error group (or abstract error group) of the nice error basis \cite{beyond1}. The error group modulo the center is called the index group of the nice error basis. 

The nice error basis provides a unitary basis for tracking error amplitudes \cite{knill1}, and generalizes the prototypical Pauli basis (the set $\{ I, X, Y, Z \}$). The group completion of the nice error basis, the error group, is required in all formulations of quantum coding in order to provide a means to construct a good code space. The center of the error group consists of multiples of identity. In quantum coding, the absolute phase (premultiplying complex number) of an error is insignificant. This is reflected in the index group, which reduces the errors modulo the phase.

We fix a nice error basis on $n$ qudits (the associated Hilbert space being $H=\mathbb{C}^{d ^{\otimes n}}$). We form the associated error group, $E$. We choose $S$ a subgroup of $E$. The group algebra, $\mathbb{C}S$ is the set of formal sums $\{ T=\sum_{s\in S} T_{s}s \}$. 

A quantum code $Q$ is a subspace of the state space $ H =\mathbb{C}^N$. We now motivate the Fourier transform approach to the construction of quantum codes. $H$ may be viewed as a module over the group algebra $\mathbb{C}S$, where $S$ as before is any subgroup of $E$. It is a well known consequence of Maschke's theorem \cite{charthry} that every module over $\mathbb{C}S$ is a direct sum of irreducible modules. Hence $H$ can be viewed as such a direct sum of irreducible modules over $\mathbb{C}S$. This motivates selection of the code space $Q$ based on the decomposition into irreducible modules of $H$. Several classes of quantum codes, including additive and Clifford codes, ultimately take advantage of this decomposition.

Modules are intimately related to the representation theory of groups. We recall the definition of a (complex, finite) group representation \cite{charthry}. If $G$ is a group, then a representation of $G$ is a homomorphism $\rho:G \rightarrow GL(W)$ for some finite-dimensional complex vector space $W$. The representation space $W$ may be viewed as a module over the group algebra $\mathbb{C}G$. An irreducible representation of a group is defined as a representation of the group with no nontrivial invariant subspaces under the induced group action. It is easy to see that an irreducible representation is equivalent to saying that the representation space is an irreducible module over the group algebra. Associated to every group representation is a character, which is a map from the group into $\mathbb{C}$. Irreducible representations afford irreducible characters. Further details of finite group representations and their characters can be found in \cite{serre, charthry}.

We now recall the definition of the generalized Fourier transforms over a finite group, and the Fourier inversion formula, following \cite{fft, serre}. We take the finite group to be a subgroup $S$ of the error group $E$ in the remainder of this paper.

Say the irreducible representations of $S$ are given by $R=\left\lbrace \rho_{i}\right\rbrace $ and let Irr($S$)$=\left\lbrace \chi_{\rho_{i}}\right\rbrace $ denote the set of irreducible characters of $S$.

Given a finite group $S$, let $\rho_{i}:S \longrightarrow GL(W_{i}), \rho_{i}\in R$ be the distinct irreducible representations of $S$, upto isomorphism, and set $n_{i} = dim (W_{i})$. Each isomorphism

\begin{equation}
\Phi:\mathbb{C}S \longrightarrow \bigoplus_{i=1}^{i=k} \mathbb{C}^{n_{i}\times n_{i} }
\end{equation}

\noindent between the group algebra $\mathbb{C}S$ and the components $\mathbb{C}^{ n_{i} \times n_{i} }$  (known as the Wedderburn components \cite{fft}), is called a Fourier transform of the group $S$. A particular isomorphism is fixed by picking a system $\left\lbrace \rho_{1},...\rho_{k}\right\rbrace$ of representatives of irreducible representations of $S$, and defining $\Phi$ as the linear extension of the mapping $s \longrightarrow \bigoplus_{i=1}^{k} \rho_{i}(s), s \in S$.

The Fourier inversion formula \cite{serre} is given by:

\begin{equation}
T_{s}=\frac{1}{|S|}\sum_{\rho_{i} \in R} n_{i}  Trace (\rho_{i}(s^{-1})a_{i})
\end{equation}

\noindent where $n_{i} = Trace(\rho_{i}(1))$ is the dimension of the representation $\rho_{i}$, and $(a_{i})_{i \in \left\lbrace 1,\ldots k \right\rbrace }$ denotes the transform domain components. We note that each $a_{i}$ lies in $\mathbb{C}^{{n_{i}}\times n_{i}}$. 

This definition of the Fourier transform naturally includes the irreducible representations of the group in question. For the case of $S$ a finite cyclic group (isomorphic to $Z_{n}$) we recover the familiar Discrete Fourier Transform. In quantum coding, the groups we deal with need not be cyclic, or even Abelian. A partial investigation of the use of the Fourier transform for quantum coding was performed in \cite{krp} for the case of $S$ Abelian and non-normal in the error group (a ``Gottesman'' subgroup as defined by them.) We perform our study using the full generality of the Fourier transform.

\section{The Fourier Inversion Formula and the Classes of Quantum Codes}

We now study the inversion formula (2) and how it is related to quantum codes. Substituting the formula (2) into the expression $T=\sum_{s\in S} T_{s}s$, we have:

\begin{equation}
T=\frac{1}{|S|}\sum_{s \in S} s\sum_{\rho_{i} \in R} n_{i} Trace (\rho_{i}(s^{-1})a_{i})
\end{equation}

We constrain $a_{i} \in \mathbb{C}^{n_{i}\times n_{i}}$ such that $a_{i}^{2} = a_{i}$. Since convolution in the group algebra maps to pointwise multiplication in the transform domain, this allows us to find operators $T \in \mathbb{C} S$ with $T^{2}=T$. These $T$, being matrices, are projectors onto subspaces of the Hilbert space $H$, and hence represent quantum codes.

We now study the projectors, or quantum codes, obtainable from Fourier inversion on different subgroups $S$ of the error group $E$, with examples. The different classes are represented in Fig. 1.

\begin{figure}
\includegraphics[scale=0.4]{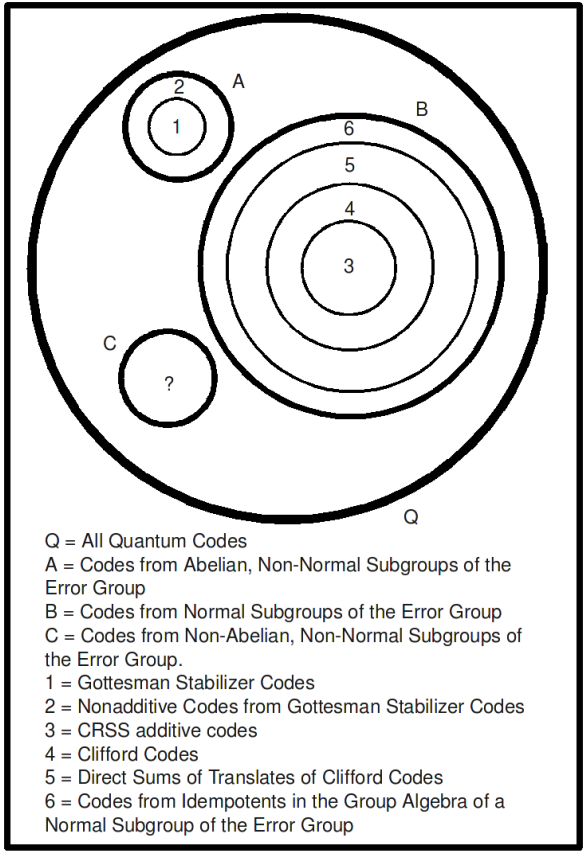} 
\caption{Venn diagram showing the classes of quantum codes obtainable by Fourier inversion}
\end{figure}

\subsection*{Class A: S an Abelian, non-normal subgroup of the error group}
Following \cite{krp}, we assume the error group to have Abelian index group. We define a ``Gottesman'' subgroup of the error group to be an Abelian subgroup of $E$, not containing any non-trivial multiple of identity $\omega I, \omega \neq 1$. In this case, $S$ is not normal in $E$. Since all the irreducible representations of the Abelian group $S$ are 1-dimensional, and there are $|S|$ of them \cite{serre}, the transform domain components are specified by a vector $A = (a_{i})_{i\in S}$. If we choose $A = (\delta_{i,1})$, we obtain from (2) the projector:

\[
T_{1} = \frac{1}{|S|} \sum_{s \in S} s 
\]

This is the projector for the stabilizer code with stabilizer group $S$. This class of codes is represented by the set ``1'' in Fig. 1. The error correcting properties are characterized by the centralizer $Z(S)$ of this group in $E$ \cite{gottes}. The $[[5,1,3]]$ perfect quantum error correcting code \cite{perfect} is an example of such a code. If we choose several $a_{i}$ equal to 1, it is possible to obtain codes that are the direct sums of translates of Image$(T_{1})$ \cite{krp}. The $((5,6,2))$ non-additive code of \cite{rains} may be constructed like this. Such codes are represented by set ``2'' in Fig 1.

\subsection*{Class B: S a Normal Subgroup of the Error Group}

For $S$ an Abelian, normal subgroup of the Pauli group for qubit systems, and a single $a_{i}$ taken as $I$, we obtain the codes of \cite{crss}. (Set``3'' in Fig. 1). In this case, not all the transform components $a_{i}$ yield nontrivial codes. We require the condition that $\chi_{\rho_{i}}(\omega I) = \omega$, which makes the character $\chi_{\rho_{i}}$ yield the eigenvalue of the code space for each operator. In the language of Clifford theory, characters with $\chi_{\rho_{i}}(\omega I) \neq \omega$ are not irreducible constituents of the restriction of the representation of the error group $E$ to $S$.

For the case of $S$ a normal (not necesarily Abelian) subgroup of the error group (not necessarily the Pauli group), and a single $a_{i}=I$ such that $P_{\chi_{\rho_{i}}}$ is not $0$, we obtain a Clifford code (Set ``4'' in Fig. 1). This is possibly a ``true'' Clifford code only if the index group of the error group is non-Abelian. If the index group of the error basis is Abelian, all the Clifford codes obtained are stabilizer codes \cite{beyond2}.

We now consider the following case: $S$ a normal, not necessarily Abelian subgroup of $E$ and more than one $a_{i}=I$.
For this case, from (2):

\[ T = \sum_{\rho_{i} \in R} \sum_{s\in S} \frac{1}{|S|}n_{i}Trace(\rho_{i}(s^{-1})a_{i})
\]

 \[
=\sum_{\rho_{i} \in R, a_{i}=I}\sum_{s \in S}\frac{1}{|S|}n_{i}\chi_{\rho_{i}}(s^{-1})s
\]

\noindent where $\chi_{\rho_{i}}$ = Trace $\rho_{i}(s)$ is the irreducible character obtained from the irreducible representation $\rho_{i}$ of $S$.

We know from representation theory that 
\[
P_{\chi_{\rho_{i}}}=\sum_{s\in S} \frac{1}{|S|}n_{i}\chi_{\rho_{i}}(s^{-1})s
\]

\noindent is the projector onto the irreducible component of the space $\mathbb{C}^{d^{\otimes n}}$ associated with $\chi_{\rho_{i}} \in Irr(S)$.

Hence,
\begin{equation}
T = \sum_{\rho_{i} \in R, a_{i}=1} P_{\chi_{\rho_{i}}}
\end{equation}

We assume without loss of generality that each $P_{\chi_{\rho_{i}}}$ is non-zero (this is equivalent to the condition that $\rho_{i}$ is an irreducible constituent of the restriction of the representation of $E$ to $S$.) The projector $T$ is a sum of projectors for Clifford codes related by translation (this is a consequence of Clifford's theorem \cite{beyond2, charthry}, which says that the $\chi_{\rho_{i}}$ are all related by conjugation.) This has been explored for Pauli groups, but not for error groups with non-Abelian index groups. We compute the error detection properties of such codes (set ``5'' in Fig. 1) in the next section, and search for an example in the error groups \cite{ueb}.

We proceed now to the most general case in class B (set ``6'' in Fig. 1). In this case, we choose $a_{i}^{2}=a_{i}$. The $a_{i}$ need not be constrained to $\left\lbrace 0, I \right\rbrace$. For example,

\[
A = \left( 
\begin{array}{cc}
1&0 \\
0&0 \\
\end{array}
\right) 
\]

and 

\[
B = \left( 
\begin{array}{cc}
0.5&-0.5 \\
-0.5&0.5 \\
\end{array}
\right) 
\]

\noindent are valid values of $a_{i}$ for a two-dimensional representation $\rho_{i}$.

Although Clifford codes have been studied, and their error correcting properties characterized, this is not true for codes from ``non-invertible idempotents in the transform domain'' (like $A$ and $B$). In \cite{Knill}, mention is made of projectors obtained from ``primitive orthogonal idempotents'' of an irreducible character. These correspond to taking a single $a_{i}$ non-zero, and this $a_{i}$ having only a single diagonal element non-zero, and equal to 1 \cite{Knill}. This results in a code-space that is strictly smaller than a Clifford code. Setting several $a_{i}$ of this format, one can obtain the span of several such spaces as a code space. Most generally, one can use the $a_{i}$ satisfying $a_{i}^2=a_{i}$ and not necessarily invertible, to obtain code spaces. However, none of these has been systematically studied in the literature.

There are two cases to be considered: when the error group has Abelian index group, and when the error group has non-Abelian index group. An error group with Abelian index group forces set ``4'' to become set ``3'', and set ``5'' to become ``direct sum of translates of additive codes'' in Fig. 1. Using the most general idempotents in the transform domain, it may be possible to obtain new codes from error groups with Abelian index groups that are neither stabilizer codes nor the sums of their translates. We feel this justifies their further study.

\subsubsection{Error Group with Abelian Index Group}
We look for codes in set ``6'' of Fig. 1 when the index group is Abelian. As no general theory exists for such codes, we performed computer search on a small (5-qubit) Pauli group. We chose our group $S$ as follows: Set $G$ to be the stabilizer group descibing the $[[5,1,3]]$ perfect quantum error correcting code of \cite{perfect}. Choose $S$ to be the centralizer of $G$ in $E$. $S$ is normal in $E$, and has 80 irreducible representations, of which sixty-four are 1-dimensional, and sixteen 2-dimensional. Only the 2-dimensional representations contribute non-zero projectors, so we formed test sets loading transform values onto a subset of them. The values came from the set:
\[
L =
\left\lbrace 0_{2}, I_{2}, P_{X+}, P_{X-}, P_{Y+}, P_{Y-}, P_{Z+}, P_{Z-}\right\rbrace 
\]

where $P_{X+}$ denotes the 1-dimensional projector onto the $+1$ eigenspace of the Pauli $X$ operator, etc. We note that using $P_{Z+}$ or $P_{Z-}$ in a single location $a_{i}$ gives the ``primitive orthogonal idempotents'' of \cite{Knill}. Unlike the 1-dimensional case, there is an infinity of possible transform components (1 or 2-dimensional projectors) yielding projectors in the group algebra. It is an open problem as to how to design the transform domain projectors in order to get a good code space.

From our computer search, we observed that many, but not all, of the projectors in the group algebra $\mathbb{C}S$ are actually contained in $\mathbb{C}G$ where $G$ is an Abelian subgroup of $S$. We denote such codes by $A(S)$, i.e. codes that are obtainable from the group algebra of an Abelian subgroup of $S$. $A(S)$ contains only stabilizer codes or sums of their translates. All the Clifford codes of $S$ are in $A(S)$. Some non-Clifford codes of $S$ also turn up in $A(S)$. For our test group $S$, and test transform components, the results are tabulated in Table 1. We note that in this test-set, the codes outside $A(S)$ have nontrivial detectable sets, but perform very poorly in terms of minimum distance. The Clifford codes of $S$ are just stabilizer codes from $Z(S)$ (\cite{beyond2}, Theorem 6), and have minimum distance 3, as expected. 

\subsubsection{Error group with non-Abelian index group}
We begin by presenting some examples of the use of (2) for obtaining projectors (or quantum codes).

\subsection*{Example using non-invertible non-zero transform values}

\[
a_{i}=A=\left( 
\begin{array}{cc}
1 & 0 \\
0 & 0 \\
\end{array}
\right) 
\]

Consider the error group on 4-level qudits, of size 32, with index group $C_{2}\times D_{8}$ \cite{clifford, ueb}. As demonstrated in \cite{clifford}, we can form the error group on one qudit and locate a normal subgroup within it that is isomorphic to a dihedral group with 16 elements. This dihedral group has four one dimensional representations and three two dimensional representations over $\mathbb{C}$ \cite{serre}. Taking the transform domain components to be $(0,0,0,0,A,0_{2},0_{2})$ where $A$ is taken as above, we obtain the projector:

\[
T=\left( 
\begin{array}{cccc}
0&0&0&0 \\
0&0.5&0&-0.5i \\
0&0&0&0 \\
0&0.5i&0&0.5 \\

\end{array}
\right) 
\]

\noindent The one-dimensional image of $T$ can detect all errors in $E$.

\subsection*{Example using only invertible non-zero transform values}
Consider the same error group $E$ on 4-level qudits, with index group $C_{2}\times D_{8}$. We consider the same dihedral group of order 16 in $E$. Taking the transform domain components to be $(0,0,0,0,I_{2},0_{2},0_{2})$ we obtain the following projector using the Fourier inversion formula:

\[
T = 
\left( 
\begin{array}{cccc}
0.5&0&0.5i&0 \\
0&0.5&0&-0.5i \\
-0.5i&0&0.5&0 \\
0&0.5i&0&0.5 \\
\end{array}
\right) 
\]

This is the same projector obtained in \cite{clifford} using the tools of Clifford theory; it is the smallest example of a Clifford code that is not a stabilizer code. We have tabulated all the projectors available from non-Abelian normal subgroups of this error group in Tables II-VIII. The transform values are taken from the same set $L=\left\lbrace 0_{2}, I_{2}, P_{X+}, P_{X-}, P_{Y+}, P_{Y-}, P_{Z+}, P_{Z-}\right\rbrace $. Hence, the tables include codes obtained using only the set $\left\lbrace 0, I\right\rbrace $, using only noninvertible non-zero transform values (like $P_{Z_{+}}$) and using a combination of invertible and non-invertible non-zero values in the transform domain. In making this table, we have considered only those transform components that contribute non-zero projectors (i.e. transform components corresponding to the irreducible constituents.) Similar tables can be generated for the other error groups in \cite{ueb}.

\section{Sums of Translates of Clifford Codes}
We assume henceforth that $a_{i}\in \left\lbrace 0, I \right\rbrace $. For this restricted case, we have seen (4) that:

\[
T = \sum_{\rho_{i} \in R, a_{i}=1} P_{\chi_{\rho_{i}}}
\]

where 
\[
P_{\chi_{\rho_{i}}}=\sum_{s\in S} \frac{1}{|S|}n_{i}\chi_{\rho_{i}}(s^{-1})s
\]

\noindent is the projector onto the irreducible component of the space $\mathbb{C}^{d^{\otimes n}}$ associated with $\chi_{\rho_{i}} \in Irr(S)$.

\noindent The code becomes a sum of translates of Clifford codes, for the $a_{i}$ chosen as above. We now treat the error detection properties of such codes. We assume without loss of generality that each $P_{\chi_{\rho_{i}}}$ is non-zero (this is equivalent to the condition that $\rho_{i}$ is an irreducible constituent of the restriction of the representation of $E$ to $S$.)

The projector $T$ is a sum of projectors for Clifford codes. By Clifford's theorem \cite{charthry}, the characters of the irreducible constitutents of the representation of S are related by conjugation.

We can rewrite (4) as $T=P_{\chi_{1}}+P_{\chi_{2}} + \ldots + P_{\chi_{t}}$. Then, Clifford's theorem says that:
\[
  \chi_{i} = \chi_{1}^{h_{i}}
\]
for some $h_{i} \in E$. Here $\chi^{h}(s)=\chi (hsh^{-1})$ for some $h \in E$. We define the set $B = \left\lbrace h_{i} \right\rbrace$. We denote the image of the space $P_{\chi_{1}}$ by $W$, and the translates of $W$ by $hW, h \in B$. The $hW$ are the images of the $P_{\chi_{1}^{h}}$. We denote the quasikernel of $hW$ by $Z(hW)$, and the inertia subgroup of $W$ by $T(W)$ \cite{beyond2}. If $W$ is obtained as the image of the projector $P_{\chi}$, we can compute the inertia group from the character $\chi$. We denote this as $T(\chi)$.

\subsection*{Error detection properties of direct sums of translates of Clifford codes}
We focus on error detection conditions for these codes. The error detection conditions for a quantum code \cite{Knill} state that an error $g$ can be detected by a code with projector $T$ if, and only if:
\[
TgT = \phi (g)T
\]

for some $\phi (g) \in \mathbb{C}$. We consider three separate cases of $g\in E$.
\subsubsection*{Case 1: $g\in \cap Z(hW)$}
We recall the definition of $T(W)$ and $Z(W)$, the inertia subgroup and quasikernel \cite{beyond2} of the Clifford code $W$:

\[
T(W)=\left\lbrace g\in E |gW \tilde{=} W \right\rbrace 
\]

$Z(W)$ is defined to be the set of elements that act on the Clifford code $Q$ by scalar multiplication:
\[
Z(W) = \left\lbrace g\in T(W) | \exists \lambda\in \mathbb{C} \forall v\in Q, gv=\lambda v \right\rbrace 
\]

In this case, $g$ acts by scalar multiplication on each space $hW$. We have:
\[
TgT = g.g^{-1}TgT
=g.(g^{-1}\sum_{h\in B} P_{\chi^{h}} g) T
\]
\[
=gTT = gT
=\sum_{h\in B} gP_{\chi^{h}}
\]
\[
=\sum_{h \in B} \lambda_{g}(h)P_{\chi^{h}} = \phi(g) \sum P_{\chi^{h}}
\]
Equating the two sides, a necessary and sufficient condition for error detection is:
\[
\forall h\in B, \lambda_{g}(h) = constant = \phi(g)
\]

It can be shown that this is a generalization of the first error correcting condition of \cite{krp} (though the subgroup $S$ taken there is a Gottesman subgroup, and not normal in the error group.)

\subsubsection*{Case 2: $g \notin S$}
In this case,
\[
TgT = g.g^{-1}TgT
\]
\[
=g.g^{-1}(\sum_{h\in B} P_{\chi^{h}} )g.T\]
\[
=g.\sum P_{\chi^{hg}}.T
\]
\[
=\phi(g).\sum_{h\in B} P_{\chi^{h}}
\]

We note that in the group algebra $\mathbb{C}E$, the LHS and RHS of the above equation have disjoint support, and hence, both sides must be zero for the error detection condition to hold.

\[
\sum_{h\in B} P_{\chi^{hg}}.T
=\sum_{h\in B} P_{\chi^{hg}} . \sum_{h' \in B} P_{\chi^{h'}}
\]

\[
=\sum_{h,h' \in B} P_{\chi^{hg}}.P_{\chi^{h^{'}}}
\]
\[
=\sum_{h,h'\in B} [\chi^{hg},\chi^{h^{'}}]P_{\chi^{h^{'}}}
\]
It is helpful to view the last step of the above equation in the transform domain. A necessary and sufficient condition for both sides of the equation to be zero is:
\[
[\chi^{hg},\chi^{h^{'}}] = 0 \forall h,h^{'} \in B
\]
\subsubsection*{Case 3:$g\in S, g \notin \cap Z(hW)$}
We note that $S$ is a subgroup of the inertia groups $T(\chi^{h}),  \forall h\in B$. We are therefore considering elements of the inertia groups that do not act by scalar multiplication on the code space. As showed in \cite{beyond2}, such errors cannot be detected by Clifford codes, hence cannot be detected by direct sums of Clifford codes either.

\section{Example of sums of translates of Clifford codes}
It is hard to find non-trivial examples of direct sums of translates of Clifford codes, as all the non-additive Clifford codes known to us consist of coding on a single qudit \cite{clifford,remarks}. Hence the number of dimensions available for packing in translates is less. We consult \cite{remarks} for the following example. It may be noted that the only other possibility of a direct sum of Clifford codes in this table is the sum of two  2-dimensional Clifford codes in a 6-dimensional space, or the sum of two 3-dimensional Clifford codes in a 9-dimensional space. Both of these yield trivial detectable sets, however.

\subsection*{Example: 2 dimensional projectors in 8 dimensional space}

We consider coding using $n=1, d=8$. We use the error basis $E$ with non-Abelian index group SmallGroup(64,10) \cite{ueb}. A normal, non-Abelian subgroup $S$ is obtained in $E$ of size 32, and yields four projectors of dimension 2 each. We take the code $C$ to be the sum of the images of two of these projectors, $P_{1}$ and $P_{2}$.

\[P_{1} = \left( 
\begin{array}{cccccccc}
1&0&0&0&0&0&0&0\\
0&0&0&0&0&0&0&0\\
0&0&0&0&0&0&0&0\\
0&0&0&0&0&0&0&0\\
0&0&0&0&1&0&0&0\\
0&0&0&0&0&0&0&0\\
0&0&0&0&0&0&0&0\\
0&0&0&0&0&0&0&0\\
\end{array}
\right) 
\]

\[P_{2} = \left( 
\begin{array}{cccccccc}
0&0&0&0&0&0&0&0\\
0&1&0&0&0&0&0&0\\
0&0&0&0&0&0&0&0\\
0&0&0&0&0&0&0&0\\
0&0&0&0&0&0&0&0\\
0&0&0&0&0&1&0&0\\
0&0&0&0&0&0&0&0\\
0&0&0&0&0&0&0&0\\
\end{array}
\right) 
\]

For the code with projector $P=P_{1} + P_{2}$ (a direct sum of Clifford codes), we computed the detectable set using the computer algebra package GAP \cite{GAP}. This is a 68-element subset of the 128-element $E$. We verified manually that the error-detection conditions previously mentioned hold true. Examples of the three classes of errors from the error group are:

\[
e_{1} = 
\left( 
\begin{array}{cccccccc}
-1 &  0 &  0 &  0 &  0 &  0 &  0 &  0 \\
 0 & -1 &  0 &  0 &  0 &  0 &  0 &  0 \\
 0 &  0 &  1 &  0 &  0 &  0 &  0 &  0 \\
 0 &  0 &  0 &  1 &  0 &  0 &  0 &  0 \\
 0 &  0 &  0 &  0 & -1 &  0 &  0 &  0 \\
 0 &  0 &  0 &  0 &  0 & -1 &  0 &  0 \\
 0 &  0 &  0 &  0 &  0 &  0 &  1 &  0 \\
 0 &  0 &  0 &  0 &  0 &  0 &  0 &  1 \\
\end{array}
\right) 
\]

\[
e_{2}=
\left( 
\begin{array}{cccccccc}
0&0 &    -1  &    0 &     0  &    0  &    0   &   0 \\
0&0&0&-1&0&0&0&0 \\
0&  -i  &    0 &     0&      0 &     0 &     0  &    0 \\
-1&      0  &    0&      0&      0&      0&      0 &    0 \\
0  &    0    &  0&      0 &     0 &     0 &     0  &   -1 \\
0   &   0    &  0&      0 &     0 &     0 & -i  &    0 \\
0    &  0    &  0&      0 &  i &     0 &     0  &    0 \\
0    &  0    &  0&      0 &     0 &  i &     0  &    0 \\
\end{array}
\right) 
\]

\[e_{3}=
\left( 
\begin{array}{cccccccc}
0&      0 &     0  &    0  &   -1 &     0 &     0 &     0 \\
0&      0 &     0  &    0  &    0 & -i &     0 &     0 \\
0&      0 &     0  &    0  &    0 &     0 &     0 &    -1 \\
0&      0 &     0  &    0  &    0 &     0 &    -1 &     0 \\
-1&      0&      0 &     0 &     0&      0&      0&      0 \\
0 &  i &     0  &    0  &    0 &     0 &     0 &     0 \\
0 &     0 &     0  &   -1  &    0 &     0 &     0 &     0 \\
0 &     0 &    -1  &    0  &    0 &     0 &     0 &     0 \\
\end{array}
\right) 
\]

We have chosen $e_{1}, e_{2}$ and $e_{3}$ corresponding to cases (1), (2) and (3) of the analysis for error detection. That is, $e_{1} \in \bigcap Z(hW)$, $e_{2} \notin S$ and $e_{3} \in S, e_{3} \notin \bigcap Z(hW)$. It is possible to verify that these errors are detectable, detectable and not detectable by the code with projector $P_{1}+P_{2}$, respectively

\section{Conclusion}
We have analyzed some well-known classes of quantum codes (stabilizer or additive codes, non-additive sums of translates of stabilizer codes, and Clifford codes) in the framework of the Fourier transform. We have analyzed the direct sums of translates of Clifford codes that also result from this framework, and produced an example using Clifford codes from an error group with non-Abelian index group, for which  detectability conditions were verified. 

A case we have not covered in this paper is: analysis of the Fourier inversion on a non-normal, non-Abelian subgroup of an error group with non-Abelian index group (class C in Fig. 1.) Such subgroups $S$ of $E$ are sporadic and the error detection properties of the code spaces obtained are not easily characterized. One reason for this is that there is no equivalent of Clifford's theorem that will tell exactly how the ambient space $\mathbb{C}^{d^{\otimes n}}$ will split as a $\mathbb{C}S$ module. For example, in the error group on one qudit with non-Abelian index group SmallGroup(36,13), a non-Abelian, non-normal subgroup $S$ exists that yields a 4-dimensional projector. However, the entire space itself being only 6-dimensional, we see that the error group no longer acts transitively on the invariant spaces of $S$, and a general characterization of error detection becomes hard.

It is an open problem to generalize the results of this paper to Operator Quantum Error Correction \cite{subsystem} and express that form of coding in the framework of the Fourier transform.

% conference papers do not normally have an appendix

% use section* for acknowledgement

% trigger a \newpage just before the given reference
% number - used to balance the columns on the last page
% adjust value as needed - may need to be readjusted if
% the document is modified later
%\IEEEtriggeratref{8}
% The "triggered" command can be changed if desired:
%\IEEEtriggercmd{\enlargethispage{-5in}}

% references section

% can use a bibliography generated by BibTeX as a .bbl file
% BibTeX documentation can be easily obtained at:
% http://www.ctan.org/tex-archive/biblio/bibtex/contrib/doc/
% The IEEEtran BibTeX style support page is at:
% http://www.michaelshell.org/tex/ieeetran/bibtex/
%\bibliographystyle{IEEEtran}
% argument is your BibTeX string definitions and bibliography database(s)
%\bibliography{IEEEabrv,../bib/paper}
%
% <OR> manually copy in the resultant .bbl file
% set second argument of \begin to the number of references
% (used to reserve space for the reference number labels box)

\newpage 
\onecolumn

\begin{table}[h]

\small

\begin{tabular}[l]{|l|c|c|c|c|c|c|c|c|}
\hline
Sl. no.&Transform&Is Clifford &In $A(S)$?&Dimension&Size of &Wt. 1& Wt. 2&Minimum\\
&Components&Code of $S$?&&&detectable set&errors(of30)&errors(of180)&distance\\
\hline
1	&	$O_{2}$	,	$P_{Z-}$	&	No	&	Yes	&	1	&	2048	&	30	&	180	&	6	\\
2	&	$O_{2}$	,	$P_{X-}$	&	No	&	Yes	&	1	&	2048	&	30	&	180	&	6	\\
3	&	$O_{2}$	,	$P_{X+}$	&	No	&	Yes	&	1	&	2048	&	30	&	180	&	6	\\
4	&	$O_{2}$	,	$P_{Y+}$	&	No	&	Yes	&	1	&	2048	&	30	&	180	&	6	\\
5	&	$O_{2}$	,	$P_{Y-}$	&	No	&	Yes	&	1	&	2048	&	30	&	180	&	6	\\
6	&	$O_{2}$	,	$P_{Z+}$	&	No	&	Yes	&	1	&	2048	&	30	&	180	&	6	\\
7	&	$O_{2}$	,	$I_{2}$	&	Yes	&	Yes	&	2	&	1952	&	30	&	180	&	3	\\
8	&	$P_{Z-}$	,	$O_{2}$	&	No	&	Yes	&	1	&	2048	&	30	&	180	&	6	\\
9	&	$P_{Z-}$	,	$P_{Z-}$	&	No	&	Yes	&	2	&	1952	&	30	&	172	&	2	\\
10	&	$P_{Z-}$	,	$P_{X-}$	&	No	&	No	&	2	&	1840	&	28	&	168	&	1	\\
11	&	$P_{Z-}$	,	$P_{X+}$	&	No	&	No	&	2	&	1840	&	28	&	168	&	1	\\
12	&	$P_{Z-}$	,	$P_{Y+}$	&	No	&	No	&	2	&	1840	&	28	&	168	&	1	\\
13	&	$P_{Z-}$	,	$P_{Y-}$	&	No	&	No	&	2	&	1840	&	28	&	168	&	1	\\
14	&	$P_{Z-}$	,	$P_{Z+}$	&	No	&	Yes	&	3	&	1808	&	30	&	172	&	1	\\
15	&	$P_{Z-}$	,	$I_{2}$	&	No	&	Yes	&	3	&	1808	&	28	&	168	&	1	\\
16	&	$P_{X-}$	,	$O_{2}$	&	No	&	Yes	&	1	&	2048	&	30	&	180	&	6	\\
17	&	$P_{X-}$	,	$P_{Z-}$	&	No	&	No	&	2	&	1840	&	28	&	168	&	1	\\
18	&	$P_{X-}$	,	$P_{X-}$	&	No	&	Yes	&	2	&	1952	&	28	&	176	&	1	\\
19	&	$P_{X-}$	,	$P_{X+}$	&	No	&	Yes	&	2	&	1952	&	30	&	172	&	2	\\
20	&	$P_{X-}$	,	$P_{Y+}$	&	No	&	No	&	2	&	1840	&	28	&	168	&	1	\\
21	&	$P_{X-}$	,	$P_{Y-}$	&	No	&	No	&	2	&	1840	&	28	&	168	&	1	\\
22	&	$P_{X-}$	,	$P_{Z+}$	&	No	&	No	&	2	&	1840	&	28	&	168	&	1	\\
23	&	$P_{X-}$	,	$I_{2}$	&	No	&	Yes	&	3	&	1808	&	28	&	168	&	1	\\
24	&	$P_{X+}$	,	$O_{2}$	&	No	&	Yes	&	1	&	2048	&	30	&	180	&	6	\\
25	&	$P_{X+}$	,	$P_{Z-}$	&	No	&	No	&	2	&	1840	&	28	&	168	&	1	\\
26	&	$P_{X+}$	,	$P_{X-}$	&	No	&	Yes	&	2	&	1952	&	30	&	172	&	2	\\
27	&	$P_{X+}$	,	$P_{X+}$	&	No	&	Yes	&	2	&	1952	&	28	&	176	&	1	\\
28	&	$P_{X+}$	,	$P_{Y+}$	&	No	&	No	&	2	&	1840	&	28	&	168	&	1	\\
29	&	$P_{X+}$	,	$P_{Y-}$	&	No	&	No	&	2	&	1840	&	28	&	168	&	1	\\
30	&	$P_{X+}$	,	$P_{Z+}$	&	No	&	No	&	2	&	1840	&	28	&	168	&	1	\\
31	&	$P_{X+}$	,	$I_{2}$	&	No	&	Yes	&	3	&	1808	&	28	&	168	&	1	\\
32	&	$P_{Y+}$	,	$O_{2}$	&	No	&	Yes	&	1	&	2048	&	30	&	180	&	6	\\
33	&	$P_{Y+}$	,	$P_{Z-}$	&	No	&	No	&	2	&	1840	&	28	&	168	&	1	\\
34	&	$P_{Y+}$	,	$P_{X-}$	&	No	&	No	&	2	&	1840	&	28	&	168	&	1	\\
35	&	$P_{Y+}$	,	$P_{X+}$	&	No	&	No	&	2	&	1840	&	28	&	168	&	1	\\
36	&	$P_{Y+}$	,	$P_{Y+}$	&	No	&	Yes	&	2	&	1952	&	30	&	172	&	2	\\
37	&	$P_{Y+}$	,	$P_{Y-}$	&	No	&	Yes	&	2	&	1952	&	28	&	176	&	1	\\
38	&	$P_{Y+}$	,	$P_{Z+}$	&	No	&	No	&	2	&	1840	&	28	&	168	&	1	\\
39	&	$P_{Y+}$	,	$I_{2}$	&	No	&	Yes	&	3	&	1808	&	28	&	168	&	1	\\
40	&	$P_{Y-}$	,	$O_{2}$	&	No	&	Yes	&	1	&	2048	&	30	&	180	&	6	\\
41	&	$P_{Y-}$	,	$P_{Z-}$	&	No	&	No	&	2	&	1840	&	28	&	168	&	1	\\
42	&	$P_{Y-}$	,	$P_{X-}$	&	No	&	No	&	2	&	1840	&	28	&	168	&	1	\\
43	&	$P_{Y-}$	,	$P_{X+}$	&	No	&	No	&	2	&	1840	&	28	&	168	&	1	\\
44	&	$P_{Y-}$	,	$P_{Y+}$	&	No	&	Yes	&	2	&	1952	&	28	&	176	&	1	\\
45	&	$P_{Y-}$	,	$P_{Y-}$	&	No	&	Yes	&	2	&	1952	&	30	&	172	&	2	\\
46	&	$P_{Y-}$	,	$P_{Z+}$	&	No	&	No	&	2	&	1840	&	28	&	168	&	1	\\
47	&	$P_{Y-}$	,	$I_{2}$	&	No	&	Yes	&	3	&	1808	&	28	&	168	&	1	\\
48	&	$P_{Z+}$	,	$O_{2}$	&	No	&	Yes	&	1	&	2048	&	30	&	180	&	6	\\
49	&	$P_{Z+}$	,	$P_{Z-}$	&	No	&	Yes	&	2	&	1952	&	28	&	176	&	1	\\
50	&	$P_{Z+}$	,	$P_{X-}$	&	No	&	No	&	2	&	1840	&	28	&	168	&	1	\\
51	&	$P_{Z+}$	,	$P_{X+}$	&	No	&	No	&	2	&	1840	&	28	&	168	&	1	\\
52	&	$P_{Z+}$	,	$P_{Y+}$	&	No	&	No	&	2	&	1840	&	28	&	168	&	1	\\
53	&	$P_{Z+}$	,	$P_{Y-}$	&	No	&	No	&	2	&	1840	&	28	&	168	&	1	\\
54	&	$P_{Z+}$	,	$P_{Z+}$	&	No	&	Yes	&	2	&	1952	&	30	&	172	&	2	\\
55	&	$P_{Z+}$	,	$I_{2}$	&	No	&	Yes	&	3	&	1808	&	28	&	168	&	1	\\
56	&	$I_{2}$	,	$O_{2}$	&	Yes	&	Yes	&	2	&	1952	&	30	&	180	&	3	\\
57	&	$I_{2}$	,	$P_{Z-}$	&	No	&	Yes	&	3	&	1808	&	28	&	168	&	1	\\
58	&	$I_{2}$	,	$P_{X-}$	&	No	&	Yes	&	3	&	1808	&	28	&	168	&	1	\\
59	&	$I_{2}$	,	$P_{X+}$	&	No	&	Yes	&	3	&	1808	&	28	&	168	&	1	\\
60	&	$I_{2}$	,	$P_{Y+}$	&	No	&	Yes	&	3	&	1808	&	28	&	168	&	1	\\
61	&	$I_{2}$	,	$P_{Y-}$	&	No	&	Yes	&	3	&	1808	&	28	&	168	&	1	\\
62	&	$I_{2}$	,	$P_{Z+}$	&	No	&	Yes	&	3	&	1808	&	28	&	168	&	1	\\
63	&	$I_{2}$	,	$I_{2}$	&	Yes	&	Yes	&	4	&	1808	&	28	&	168	&	1	\\

\hline
\end{tabular}
\caption{Computer search results for codes from idempotents in the transform domain,for $S$}

\end{table}

\twocolumn

\begin{table}[h]
\small

Error group Id = [ 32, 6 ] \\
Index group Id = [ 16, 3 ] \\

\begin{tabular}{|c|c|c|c|c|}
\hline
Tx domain&Dimension&Size of \\
components&&detectable set\\
\hline
$I_{2}$&4&2\\
$P_{Z_{+}}$&2&10\\
$P_{Z_{-}}$&2&10\\
$P_{X_{+}}$&2&10\\
$P_{X_{-}}$&2&10\\
$P_{Y_{+}}$&2&20\\
$P_{Y_{-}}$&2&20\\
\hline
\end{tabular}
\caption{ Projectors from idempotents for subgroup with ID [8,3]}
\end{table}

\begin{table}[h]
\small

\begin{tabular}{|c|c|c|}
\hline
Tx domain&Dimension&Size of \\
components&&detectable set\\
\hline
 $I_{2}$&4&2\\
 $P_{Z_{+}}$&2&10\\
 $P_{Z_{-}}$&2&10\\
 $P_{X_{+}}$&2&10\\
 $P_{X_{-}}$&2&10\\
 $P_{Y_{+}}$&2&20\\
 $P_{Y_{-}}$&2&20\\
\hline
\end{tabular}
\caption{ Projectors from idempotents for subgroup with ID [8,4]}
\end{table}
\begin{table}
\small

\begin{tabular}{|c|c|c|c|}
\hline
Tx domain&Dimension&Size of \\
components&&detectable set\\
\hline
 $O_{2}$,$I_{2}$&2&12\\
 $O_{2}$,$P_{Z_{+}}$&1&32\\
 $O_{2}$,$P_{Z_{-}}$&1&32\\
 $O_{2}$,$P_{X_{+}}$&1&32\\
 $O_{2}$,$P_{X_{-}}$&1&32\\
 $O_{2}$,$P_{Y_{+}}$&1&32\\
 $O_{2}$,$P_{Y_{-}}$&1&32\\
 $I_{2}$,$O_{2}$&2&12\\
 $I_{2}$,$I_{2}$&4&2\\
 $I_{2}$,$P_{Z_{+}}$&3&2\\
 $I_{2}$,$P_{Z_{-}}$&3&2\\
 $I_{2}$,$P_{X_{+}}$&3&2\\
 $I_{2}$,$P_{X_{-}}$&3&2\\
 $I_{2}$,$P_{Y_{+}}$&3&2\\
 $I_{2}$,$P_{Y_{-}}$&3&2\\
 $P_{Z_{+}}$,$O_{2}$&1&32\\
 $P_{Z_{+}}$,$I_{2}$&3&2\\
 $P_{Z_{+}}$,$P_{Z_{+}}$&2&20\\
 $P_{Z_{+}}$,$P_{Z_{-}}$&2&12\\
 $P_{Z_{+}}$,$P_{X_{+}}$&2&2\\
 $P_{Z_{+}}$,$P_{X_{-}}$&2&2\\
 $P_{Z_{+}}$,$P_{Y_{+}}$&2&2\\
 $P_{Z_{+}}$,$P_{Y_{-}}$&2&2\\
 $P_{Z_{-}}$,$O_{2}$&1&32\\
 $P_{Z_{-}}$,$I_{2}$&3&2\\
 $P_{Z_{-}}$,$P_{Z_{+}}$&2&12\\
 $P_{Z_{-}}$,$P_{Z_{-}}$&2&20\\
 $P_{Z_{-}}$,$P_{X_{+}}$&2&2\\
 $P_{Z_{-}}$,$P_{X_{-}}$&2&2\\
 $P_{Z_{-}}$,$P_{Y_{+}}$&2&2\\
 $P_{Z_{-}}$,$P_{Y_{-}}$&2&2\\
 $P_{X_{+}}$,$O_{2}$&1&32\\
 $P_{X_{+}}$,$I_{2}$&3&2\\
 $P_{X_{+}}$,$P_{Z_{+}}$&2&2\\
 $P_{X_{+}}$,$P_{Z_{-}}$&2&2\\
 $P_{X_{+}}$,$P_{X_{+}}$&2&6\\
 $P_{X_{+}}$,$P_{X_{-}}$&2&6\\
 $P_{X_{+}}$,$P_{Y_{+}}$&2&10\\
 $P_{X_{+}}$,$P_{Y_{-}}$&2&10\\
 $P_{X_{-}}$,$O_{2}$&1&32\\
 $P_{X_{-}}$,$I_{2}$&3&2\\
 $P_{X_{-}}$,$P_{Z_{+}}$&2&2\\
 $P_{X_{-}}$,$P_{Z_{-}}$&2&2\\
 $P_{X_{-}}$,$P_{X_{+}}$&2&6\\
 $P_{X_{-}}$,$P_{X_{-}}$&2&6\\
 $P_{X_{-}}$,$P_{Y_{+}}$&2&10\\
 $P_{X_{-}}$,$P_{Y_{-}}$&2&10\\
 $P_{Y_{+}}$,$O_{2}$&1&32\\
 $P_{Y_{+}}$,$I_{2}$&3&2\\
 $P_{Y_{+}}$,$P_{Z_{+}}$&2&2\\
 $P_{Y_{+}}$,$P_{Z_{-}}$&2&2\\
 $P_{Y_{+}}$,$P_{X_{+}}$&2&10\\
 $P_{Y_{+}}$,$P_{X_{-}}$&2&10\\
 $P_{Y_{+}}$,$P_{Y_{+}}$&2&6\\
 $P_{Y_{+}}$,$P_{Y_{-}}$&2&6\\
 $P_{Y_{-}}$,$O_{2}$&1&32\\
 $P_{Y_{-}}$,$I_{2}$&3&2\\
 $P_{Y_{-}}$,$P_{Z_{+}}$&2&2\\
 $P_{Y_{-}}$,$P_{Z_{-}}$&2&2\\
 $P_{Y_{-}}$,$P_{X_{+}}$&2&10\\
 $P_{Y_{-}}$,$P_{X_{-}}$&2&10\\
 $P_{Y_{-}}$,$P_{Y_{+}}$&2&6\\
 $P_{Y_{-}}$,$P_{Y_{-}}$&2&6\\
\hline
\end{tabular}
\caption{ Projectors from idempotents for subgroup with ID [16,13]}
\end{table}

\begin{table}
\small

\begin{tabular}{|c|c|c|c}
\hline
Tx domain&Dimension&Size of \\
components&&detectable set\\
\hline
 $O_{2}$,$I_{2}$&2&12\\
 $O_{2}$,$P_{Z_{+}}$&1&32\\
 $O_{2}$,$P_{Z_{-}}$&1&32\\
 $O_{2}$,$P_{X_{+}}$&1&32\\
 $O_{2}$,$P_{X_{-}}$&1&32\\
 $O_{2}$,$P_{Y_{+}}$&1&32\\
 $O_{2}$,$P_{Y_{-}}$&1&32\\
 $I_{2}$,$O_{2}$&2&12\\
 $I_{2}$,$I_{2}$&4&2\\
 $I_{2}$,$P_{Z_{+}}$&3&2\\
 $I_{2}$,$P_{Z_{-}}$&3&2\\
 $I_{2}$,$P_{X_{+}}$&3&2\\
 $I_{2}$,$P_{X_{-}}$&3&2\\
 $I_{2}$,$P_{Y_{+}}$&3&2\\
 $I_{2}$,$P_{Y_{-}}$&3&2\\
 $P_{Z_{+}}$,$O_{2}$&1&32\\
 $P_{Z_{+}}$,$I_{2}$&3&2\\
 $P_{Z_{+}}$,$P_{Z_{+}}$&2&2\\
 $P_{Z_{+}}$,$P_{Z_{-}}$&2&2\\
 $P_{Z_{+}}$,$P_{X_{+}}$&2&2\\
 $P_{Z_{+}}$,$P_{X_{-}}$&2&2\\
 $P_{Z_{+}}$,$P_{Y_{+}}$&2&2\\
 $P_{Z_{+}}$,$P_{Y_{-}}$&2&2\\
 $P_{Z_{-}}$,$O_{2}$&1&32\\
 $P_{Z_{-}}$,$I_{2}$&3&2\\
 $P_{Z_{-}}$,$P_{Z_{+}}$&2&2\\
 $P_{Z_{-}}$,$P_{Z_{-}}$&2&2\\
 $P_{Z_{-}}$,$P_{X_{+}}$&2&2\\
 $P_{Z_{-}}$,$P_{X_{-}}$&2&2\\
 $P_{Z_{-}}$,$P_{Y_{+}}$&2&2\\
 $P_{Z_{-}}$,$P_{Y_{-}}$&2&2\\
 $P_{X_{+}}$,$O_{2}$&1&32\\
 $P_{X_{+}}$,$I_{2}$&3&2\\
 $P_{X_{+}}$,$P_{Z_{+}}$&2&2\\
 $P_{X_{+}}$,$P_{Z_{-}}$&2&2\\
 $P_{X_{+}}$,$P_{X_{+}}$&2&2\\
 $P_{X_{+}}$,$P_{X_{-}}$&2&2\\
 $P_{X_{+}}$,$P_{Y_{+}}$&2&2\\
 $P_{X_{+}}$,$P_{Y_{-}}$&2&2\\
 $P_{X_{-}}$,$O_{2}$&1&32\\
 $P_{X_{-}}$,$I_{2}$&3&2\\
 $P_{X_{-}}$,$P_{Z_{+}}$&2&2\\
 $P_{X_{-}}$,$P_{Z_{-}}$&2&2\\
 $P_{X_{-}}$,$P_{X_{+}}$&2&2\\
 $P_{X_{-}}$,$P_{X_{-}}$&2&2\\
 $P_{X_{-}}$,$P_{Y_{+}}$&2&2\\
 $P_{X_{-}}$,$P_{Y_{-}}$&2&2\\
 $P_{Y_{+}}$,$O_{2}$&1&32\\
 $P_{Y_{+}}$,$I_{2}$&3&2\\
 $P_{Y_{+}}$,$P_{Z_{+}}$&2&6\\
 $P_{Y_{+}}$,$P_{Z_{-}}$&2&6\\
 $P_{Y_{+}}$,$P_{X_{+}}$&2&6\\
 $P_{Y_{+}}$,$P_{X_{-}}$&2&6\\
 $P_{Y_{+}}$,$P_{Y_{+}}$&2&2\\
 $P_{Y_{+}}$,$P_{Y_{-}}$&2&2\\
 $P_{Y_{-}}$,$O_{2}$&1&32\\
 $P_{Y_{-}}$,$I_{2}$&3&2\\
 $P_{Y_{-}}$,$P_{Z_{+}}$&2&6\\
 $P_{Y_{-}}$,$P_{Z_{-}}$&2&6\\
 $P_{Y_{-}}$,$P_{X_{+}}$&2&6\\
 $P_{Y_{-}}$,$P_{X_{-}}$&2&6\\
 $P_{Y_{-}}$,$P_{Y_{+}}$&2&2\\
 $P_{Y_{-}}$,$P_{Y_{-}}$&2&2\\
 \hline
\end{tabular}
\caption{ Projectors from idempotents for subgroup with ID [16,8]}
\end{table}

\begin{table}
\small

\begin{tabular}{|c|c|c|c|}
\hline
Tx domain&Dimension&Size of \\
components&&detectable set\\
\hline
 $O_{2}$,$I_{2}$&2&12\\
 $O_{2}$,$P_{Z_{+}}$&1&32\\
 $O_{2}$,$P_{Z_{-}}$&1&32\\
 $O_{2}$,$P_{X_{+}}$&1&32\\
 $O_{2}$,$P_{X_{-}}$&1&32\\
 $O_{2}$,$P_{Y_{+}}$&1&32\\
 $O_{2}$,$P_{Y_{-}}$&1&32\\
 $I_{2}$,$O_{2}$&2&12\\
 $I_{2}$,$I_{2}$&4&2\\
 $I_{2}$,$P_{Z_{+}}$&3&2\\
 $I_{2}$,$P_{Z_{-}}$&3&2\\
 $I_{2}$,$P_{X_{+}}$&3&2\\
 $I_{2}$,$P_{X_{-}}$&3&2\\
 $I_{2}$,$P_{Y_{+}}$&3&2\\
 $I_{2}$,$P_{Y_{-}}$&3&2\\
 $P_{Z_{+}}$,$O_{2}$&1&32\\
 $P_{Z_{+}}$,$I_{2}$&3&2\\
 $P_{Z_{+}}$,$P_{Z_{+}}$&2&6\\
 $P_{Z_{+}}$,$P_{Z_{-}}$&2&6\\
 $P_{Z_{+}}$,$P_{X_{+}}$&2&10\\
 $P_{Z_{+}}$,$P_{X_{-}}$&2&10\\
 $P_{Z_{+}}$,$P_{Y_{+}}$&2&2\\
 $P_{Z_{+}}$,$P_{Y_{-}}$&2&2\\
 $P_{Z_{-}}$,$O_{2}$&1&32\\
 $P_{Z_{-}}$,$I_{2}$&3&2\\
 $P_{Z_{-}}$,$P_{Z_{+}}$&2&6\\
 $P_{Z_{-}}$,$P_{Z_{-}}$&2&6\\
 $P_{Z_{-}}$,$P_{X_{+}}$&2&10\\
 $P_{Z_{-}}$,$P_{X_{-}}$&2&10\\
 $P_{Z_{-}}$,$P_{Y_{+}}$&2&2\\
 $P_{Z_{-}}$,$P_{Y_{-}}$&2&2\\
 $P_{X_{+}}$,$O_{2}$&1&32\\
 $P_{X_{+}}$,$I_{2}$&3&2\\
 $P_{X_{+}}$,$P_{Z_{+}}$&2&10\\
 $P_{X_{+}}$,$P_{Z_{-}}$&2&10\\
 $P_{X_{+}}$,$P_{X_{+}}$&2&6\\
 $P_{X_{+}}$,$P_{X_{-}}$&2&6\\
 $P_{X_{+}}$,$P_{Y_{+}}$&2&2\\
 $P_{X_{+}}$,$P_{Y_{-}}$&2&2\\
 $P_{X_{-}}$,$O_{2}$&1&32\\
 $P_{X_{-}}$,$I_{2}$&3&2\\
 $P_{X_{-}}$,$P_{Z_{+}}$&2&10\\
 $P_{X_{-}}$,$P_{Z_{-}}$&2&10\\
 $P_{X_{-}}$,$P_{X_{+}}$&2&6\\
 $P_{X_{-}}$,$P_{X_{-}}$&2&6\\
 $P_{X_{-}}$,$P_{Y_{+}}$&2&2\\
 $P_{X_{-}}$,$P_{Y_{-}}$&2&2\\
 $P_{Y_{+}}$,$O_{2}$&1&32\\
 $P_{Y_{+}}$,$I_{2}$&3&2\\
 $P_{Y_{+}}$,$P_{Z_{+}}$&2&2\\
 $P_{Y_{+}}$,$P_{Z_{-}}$&2&2\\
 $P_{Y_{+}}$,$P_{X_{+}}$&2&2\\
 $P_{Y_{+}}$,$P_{X_{-}}$&2&2\\
 $P_{Y_{+}}$,$P_{Y_{+}}$&2&20\\
 $P_{Y_{+}}$,$P_{Y_{-}}$&2&12\\
 $P_{Y_{-}}$,$O_{2}$&1&32\\
 $P_{Y_{-}}$,$I_{2}$&3&2\\
 $P_{Y_{-}}$,$P_{Z_{+}}$&2&2\\
 $P_{Y_{-}}$,$P_{Z_{-}}$&2&2\\
 $P_{Y_{-}}$,$P_{X_{+}}$&2&2\\
 $P_{Y_{-}}$,$P_{X_{-}}$&2&2\\
 $P_{Y_{-}}$,$P_{Y_{+}}$&2&12\\
 $P_{Y_{-}}$,$P_{Y_{-}}$&2&20\\
 \hline
\end{tabular}
\caption{ Projectors from idempotents for subgroup with ID [16,7]}
\end{table}

\begin{table}
\small

\begin{tabular}{|c|c|c|c|}
\hline
Tx domain&Dimension&Size of \\
components&&detectable set\\
\hline
 $O_{2}$,$I_{2}$&2&20\\
 $O_{2}$,$P_{Z_{+}}$&1&32\\
 $O_{2}$,$P_{Z_{-}}$&1&32\\
 $O_{2}$,$P_{X_{+}}$&1&32\\
 $O_{2}$,$P_{X_{-}}$&1&32\\
 $O_{2}$,$P_{Y_{+}}$&1&32\\
 $O_{2}$,$P_{Y_{-}}$&1&32\\
 $I_{2}$,$O_{2}$&2&20\\
 $I_{2}$,$I_{2}$&4&2\\
 $I_{2}$,$P_{Z_{+}}$&3&2\\
 $I_{2}$,$P_{Z_{-}}$&3&2\\
 $I_{2}$,$P_{X_{+}}$&3&2\\
 $I_{2}$,$P_{X_{-}}$&3&2\\
 $I_{2}$,$P_{Y_{+}}$&3&2\\
 $I_{2}$,$P_{Y_{-}}$&3&2\\
 $P_{Z_{+}}$,$O_{2}$&1&32\\
 $P_{Z_{+}}$,$I_{2}$&3&2\\
 $P_{Z_{+}}$,$P_{Z_{+}}$&2&12\\
 $P_{Z_{+}}$,$P_{Z_{-}}$&2&12\\
 $P_{Z_{+}}$,$P_{X_{+}}$&2&2\\
 $P_{Z_{+}}$,$P_{X_{-}}$&2&2\\
 $P_{Z_{+}}$,$P_{Y_{+}}$&2&2\\
 $P_{Z_{+}}$,$P_{Y_{-}}$&2&2\\
 $P_{Z_{-}}$,$O_{2}$&1&32\\
 $P_{Z_{-}}$,$I_{2}$&3&2\\
 $P_{Z_{-}}$,$P_{Z_{+}}$&2&12\\
 $P_{Z_{-}}$,$P_{Z_{-}}$&2&12\\
 $P_{Z_{-}}$,$P_{X_{+}}$&2&2\\
 $P_{Z_{-}}$,$P_{X_{-}}$&2&2\\
 $P_{Z_{-}}$,$P_{Y_{+}}$&2&2\\
 $P_{Z_{-}}$,$P_{Y_{-}}$&2&2\\
 $P_{X_{+}}$,$O_{2}$&1&32\\
 $P_{X_{+}}$,$I_{2}$&3&2\\
 $P_{X_{+}}$,$P_{Z_{+}}$&2&2\\
 $P_{X_{+}}$,$P_{Z_{-}}$&2&2\\
 $P_{X_{+}}$,$P_{X_{+}}$&2&10\\
 $P_{X_{+}}$,$P_{X_{-}}$&2&10\\
 $P_{X_{+}}$,$P_{Y_{+}}$&2&10\\
 $P_{X_{+}}$,$P_{Y_{-}}$&2&10\\
 $P_{X_{-}}$,$O_{2}$&1&32\\
 $P_{X_{-}}$,$I_{2}$&3&2\\
 $P_{X_{-}}$,$P_{Z_{+}}$&2&2\\
 $P_{X_{-}}$,$P_{Z_{-}}$&2&2\\
 $P_{X_{-}}$,$P_{X_{+}}$&2&10\\
 $P_{X_{-}}$,$P_{X_{-}}$&2&10\\
 $P_{X_{-}}$,$P_{Y_{+}}$&2&10\\
 $P_{X_{-}}$,$P_{Y_{-}}$&2&10\\
 $P_{Y_{+}}$,$O_{2}$&1&32\\
 $P_{Y_{+}}$,$I_{2}$&3&2\\
 $P_{Y_{+}}$,$P_{Z_{+}}$&2&2\\
 $P_{Y_{+}}$,$P_{Z_{-}}$&2&2\\
 $P_{Y_{+}}$,$P_{X_{+}}$&2&10\\
 $P_{Y_{+}}$,$P_{X_{-}}$&2&10\\
 $P_{Y_{+}}$,$P_{Y_{+}}$&2&10\\
 $P_{Y_{+}}$,$P_{Y_{-}}$&2&10\\
 $P_{Y_{-}}$,$O_{2}$&1&32\\
 $P_{Y_{-}}$,$I_{2}$&3&2\\
 $P_{Y_{-}}$,$P_{Z_{+}}$&2&2\\
 $P_{Y_{-}}$,$P_{Z_{-}}$&2&2\\
 $P_{Y_{-}}$,$P_{X_{+}}$&2&10\\
 $P_{Y_{-}}$,$P_{X_{-}}$&2&10\\
 $P_{Y_{-}}$,$P_{Y_{+}}$&2&10\\
 $P_{Y_{-}}$,$P_{Y_{-}}$&2&10\\
 \hline
\end{tabular}
\caption{ Projectors from idempotents for subgroup with ID [16,6]}
\end{table}

\begin{table}
\small

\begin{tabular}{|c|c|c|c|}
\hline
Tx domain&Dimension&Size of \\
components&&detectable set\\
\hline
 $O_{2}$,$I_{2}$&2&12\\
 $O_{2}$,$P_{Z_{+}}$&1&32\\
 $O_{2}$,$P_{Z_{-}}$&1&32\\
 $O_{2}$,$P_{X_{+}}$&1&32\\
 $O_{2}$,$P_{X_{-}}$&1&32\\
 $O_{2}$,$P_{Y_{+}}$&1&32\\
 $O_{2}$,$P_{Y_{-}}$&1&32\\
 $I_{2}$,$O_{2}$&2&12\\
 $I_{2}$,$I_{2}$&4&2\\
 $I_{2}$,$P_{Z_{+}}$&3&2\\
 $I_{2}$,$P_{Z_{-}}$&3&2\\
 $I_{2}$,$P_{X_{+}}$&3&2\\
 $I_{2}$,$P_{X_{-}}$&3&2\\
 $I_{2}$,$P_{Y_{+}}$&3&2\\
 $I_{2}$,$P_{Y_{-}}$&3&2\\
 $P_{Z_{+}}$,$O_{2}$&1&32\\
 $P_{Z_{+}}$,$I_{2}$&3&2\\
 $P_{Z_{+}}$,$P_{Z_{+}}$&2&10\\
 $P_{Z_{+}}$,$P_{Z_{-}}$&2&10\\
 $P_{Z_{+}}$,$P_{X_{+}}$&2&6\\
 $P_{Z_{+}}$,$P_{X_{-}}$&2&6\\
 $P_{Z_{+}}$,$P_{Y_{+}}$&2&2\\
 $P_{Z_{+}}$,$P_{Y_{-}}$&2&2\\
 $P_{Z_{-}}$,$O_{2}$&1&32\\
 $P_{Z_{-}}$,$I_{2}$&3&2\\
 $P_{Z_{-}}$,$P_{Z_{+}}$&2&10\\
 $P_{Z_{-}}$,$P_{Z_{-}}$&2&10\\
 $P_{Z_{-}}$,$P_{X_{+}}$&2&6\\
 $P_{Z_{-}}$,$P_{X_{-}}$&2&6\\
 $P_{Z_{-}}$,$P_{Y_{+}}$&2&2\\
 $P_{Z_{-}}$,$P_{Y_{-}}$&2&2\\
 $P_{X_{+}}$,$O_{2}$&1&32\\
 $P_{X_{+}}$,$I_{2}$&3&2\\
 $P_{X_{+}}$,$P_{Z_{+}}$&2&6\\
 $P_{X_{+}}$,$P_{Z_{-}}$&2&6\\
 $P_{X_{+}}$,$P_{X_{+}}$&2&10\\
 $P_{X_{+}}$,$P_{X_{-}}$&2&10\\
 $P_{X_{+}}$,$P_{Y_{+}}$&2&2\\
 $P_{X_{+}}$,$P_{Y_{-}}$&2&2\\
 $P_{X_{-}}$,$O_{2}$&1&32\\
 $P_{X_{-}}$,$I_{2}$&3&2\\
 $P_{X_{-}}$,$P_{Z_{+}}$&2&6\\
 $P_{X_{-}}$,$P_{Z_{-}}$&2&6\\
 $P_{X_{-}}$,$P_{X_{+}}$&2&10\\
 $P_{X_{-}}$,$P_{X_{-}}$&2&10\\
 $P_{X_{-}}$,$P_{Y_{+}}$&2&2\\
 $P_{X_{-}}$,$P_{Y_{-}}$&2&2\\
 $P_{Y_{+}}$,$O_{2}$&1&32\\
 $P_{Y_{+}}$,$I_{2}$&3&2\\
 $P_{Y_{+}}$,$P_{Z_{+}}$&2&2\\
 $P_{Y_{+}}$,$P_{Z_{-}}$&2&2\\
 $P_{Y_{+}}$,$P_{X_{+}}$&2&2\\
 $P_{Y_{+}}$,$P_{X_{-}}$&2&2\\
 $P_{Y_{+}}$,$P_{Y_{+}}$&2&12\\
 $P_{Y_{+}}$,$P_{Y_{-}}$&2&20\\
 $P_{Y_{-}}$,$O_{2}$&1&32\\
 $P_{Y_{-}}$,$I_{2}$&3&2\\
 $P_{Y_{-}}$,$P_{Z_{+}}$&2&2\\
 $P_{Y_{-}}$,$P_{Z_{-}}$&2&2\\
 $P_{Y_{-}}$,$P_{X_{+}}$&2&2\\
 $P_{Y_{-}}$,$P_{X_{-}}$&2&2\\
 $P_{Y_{-}}$,$P_{Y_{+}}$&2&20\\
 $P_{Y_{-}}$,$P_{Y_{-}}$&2&12\\
 \hline
\end{tabular}
\caption{ Projectors from idempotents for subgroup with ID [16,11]}
\end{table}

% that's all folks
\end{document}